\documentclass{ecai}
\usepackage[utf8]{inputenc}
\usepackage{times}
\usepackage{graphicx}
\usepackage{latexsym}

\usepackage{amsmath}
\usepackage{amssymb}
\usepackage{bm}

\usepackage[super]{nth}

\DeclareMathOperator{\AppS}{AppS}
\DeclareMathOperator{\MobS}{MobS}

\usepackage[nameinlink,noabbrev]{cleveref}

\usepackage{color}
\definecolor{Orange}{rgb}{0.9,0.5,0}
\definecolor{NavyBlue}{rgb}{0.1, 0.4, 0.8}
\definecolor{Magenta}{rgb}{0.8, 0.1, 0.6}
\definecolor{Yellow}{rgb}{0.9, 0.8, 0.4}
\definecolor{Red}{rgb}{1, 0,0}

\begin{document}

\title{Understanding individual behaviour: from virtual to physical patterns}

\author{Marco De Nadai \institute{Fondazione Bruno Kessler,
Italy, email: denadai@fbk.eu}
\ \thanks{Work done while at Vodafone Research, London (UK)}
\and Bruno Lepri \institute{Fondazione Bruno Kessler,
Italy, email: lepri@fbk.eu} \and Nuria Oliver\institute{Data-Pop Alliance} \ \thanks{Work done while at Vodafone Research, London (UK)}}

\maketitle
\bibliographystyle{ecai}

\begin{abstract}
  As ``Big Data" has become pervasive, an increasing amount of research has connected the dots between human behaviour in the offline and online worlds. Consequently, researchers have exploited these new findings to create models that better predict different aspects of human life and recommend future behaviour. 
  To date, however, we do not yet fully understand the similarities and differences of human behaviour in these virtual and physical worlds.
  Here, we analyse and discuss the mobility and application usage of 400,000 individuals over eight months. We find an astonishing similarity between people's mobility in the physical space and how they move from app to app in smartphones. Our data shows that individuals use and visit a finite number of apps and places, but they keep exploring over time. In particular, two distinct profiles of individuals emerge: those that keep changing places and services, and those that are stable over time, named as ``explorers" and ``keepers". 
  We see these findings as crucial to enrich a discussion for the potentials and the challenges of building human-centric AI systems, which might leverage recent results in Computational Social Science. 
\end{abstract}

\section{Introduction}
The rapid digitalisation of human life is increasingly blurring the boundaries between the digital and the real worlds. Internet and mobile devices have indeed become the preferred mean for many human activities such as social interactions and entertainment. 
Do people exhibit similar behaviours in the digital and physical  worlds? 

Recent seminal work has highlighted how similar mechanisms seem to govern various aspects of human activities. For example, people exhibit a finite number of friends and favourite places, which are most probably connected to the inherent constraints of human cognition~\cite{alessandretti2018evidence, miritello2013limited}. 
Similarly, some people tend to explore and change favourite places over time, as they do with friendships, while others tend always to maintain stable their behaviour~\cite{miritello2013limited, pappalardo2015returners}. 
This \emph{exploration} and \emph{exploitation} behaviour is well known in many disciplines such as computer science, where it is used in reinforcement learning or recommendation algorithms to increase the diversity of the suggestions~\cite{li2010contextual, tang2014ensemble}.
Although most of the literature has explored many similarities between people's activities, little is known about the similarities of human behaviour between the digital and physical worlds.

Here, we highlight recent research towards a direct comparison between the use of mobile applications and mobility~\cite{de2019strategies}. We explore the activity of 400,000 individuals for eight months and collect the foreground application activity as well as the visited GPS locations. All data is collected with the fully informed consent, pseudonymised and in agreement with existing data privacy and data protection regulations. 
Through this work, we found that the very same mechanisms that govern human mobility seem to play the same role in app usage.

\section{Mobility is statistically similar to app usage}
We analysed the application usage of more than 400,000 individuals whose data was available for at least 80\% of the hours in the eight month period.
GPS locations are collected through either the GPS or the WiFi connection, while application usage is collected in an Android application that tracks foreground usage. To have a uniform analysis, we consider only the applications present in the Google Play store.
We begin by describing, for the first time, the statistics of foreground app usage at scale.

We find that the app usage is well described by a truncated power-law, where the frequency $f$ of the $k^{th}$ most visited location is well approximated by: $f_k \sim (k + k_0)^{-\alpha} exp(-k/c) $, with exponent $\alpha = 1.19 \pm 0.01$, $k_0 = 1.14 \pm 0.07$ and a cut off value $c = 8.32 \pm 0.75$, fitted through~\cite{alstott2014powerlaw}. 
Similarly, mobility is well described by a power law distribution $f_k \sim k^{-\alpha}$ with $\alpha = 1.27 \pm 0.01$, which is in line with results in literature~\cite{song2010modelling} ($\alpha = 1.2 \pm 0.1$). 
Thus, the time spent on phones is mostly focused on a few apps, although users possess, on average, at least 26 applications. In mobility, the time spent on places is mainly focused on a few places. 

However, human behaviour changes over time. Let the total number of apps be $L(t)_{apps} \propto L_0 t^{\gamma_1}$, where $t$ is the time, $L_0$ is a normalisation constant, and $\gamma_1$ a growing coefficient. Similarly, we define the number of locations is $L(t)_{mob} \propto L_0 t^{\gamma_2}$.
We find that $\gamma_1=0.41$ and $\gamma_2 = 0.64$, which means that people increase the number of apps and places over time but also that people explore in the physical world at a faster pace than in the digital world, even if there are no apparent physical constraints in it. 
While app usage and mobility is skewed towards a few apps and places, people keep exploring new items.
To explain this seemingly contradicting result, we resort the concept of ``activity space"~\cite{alessandretti2018evidence}.

\section{App capacity is conserved}
In mobility, the ``activity space" is defined as the set of stop locations an individual $i$ visits at least twice and where (s)he spends on average more than 10 min per week over a time-window $t$: $MobS_i(t) = [\mathfrak{l}_1, \mathfrak{l}_2, \ldots, \mathfrak{l}_n]$. Similarly, we define the \emph{app space} as the set of applications $AppS_i (t) = [\mathfrak{a}_1, \mathfrak{a}_2, \ldots, \mathfrak{a}_n]$ that an user $i$ used at least twice in a time window $t$. Following the literature~\cite{alessandretti2018evidence} we set the time window to be 20 weeks long. This definition allows to observe the user's apps over time.

Let $C_i^{\textit{apps}}(t) = |\AppS_i (t)|$ be the number of distinct applications a user $i$ uses over a time window $t$, and $C_i^{\textit{mob}}(t) = |\MobS_i (t)|$ the similar definition for mobility. 
When we observe the evolution of the capacity over time, we find that this number is almost constant for all the users, which is around 27 for apps and 25 for mobility.
Moreover, the gain between the added and deleted apps \emph{app gain} $G_i^{\textit{apps}}(t) = A_i^{\textit{apps}}(t) - D_i^{\textit{apps}}(t)$ is on average constant. When we analyse its coefficient of variation (CV) ($\left|\left<G_i\right>^{\textit{apps}}\right| / \sigma_{G^{\textit{apps}}_i}$), we find that $97.3\%$ of the people in our data have $\left|\left<G_i\right>^{\textit{apps}}\right| / \sigma_{G_i^{\textit{apps}}} \leq 1$, thus exhibiting a conserved \emph{app capacity} ($97.5\%$ for mobility). 

However, app and mobility capacities might be constant because of time constraints. People do not (unfortunately) have infinite time. Thus, they might use (visit) a finite number of apps (places) just because of time. Yet, this result is strong against different types of randomisation~\cite{de2019strategies}, which show that the constant capacity is not a consequence of time constraints.

Capacity is constant in both the app and mobility spaces, but people keep exploring new apps and places. Thus, when an app (place) enters in the app (activity) space, another app goes out from it. 
This seemly surprising result is not new in literature. Miritello \emph{et al.}~\cite{miritello2013limited} indeed found that people behave similarly in friendships, while Alessandretti \emph{et al.} found similar results in mobility~\cite{alessandretti2018evidence}.

\section{People are either explorers or keepers with apps}
While the growth exponent of apps is $\gamma_1=0.41$ for the entire population, an average user with average capacity $\left<C_i\right> = 80$ discovers new apps at a much larger rate ($\gamma_1=0.53$). Similarly, individuals with $\left<C_i\right> = 20$ discover new locations at a much slower pace than people with $\left<C_i\right> = 80$ ($\gamma_2=0.58$ vs $\gamma_2=0.78$).
Thus, let $R^{\textit{apps}}_{i} = \left<A_i\right>^{\textit{apps}} / \left<C_i\right>^{\textit{apps}}$ be the ratio between the added apps and the capacity of a user $i$. 
We define application \emph{explorers} to be those users with $R_{i} \gg \beta$ and application \emph{keepers} to be those users with $R_{i} \ll \beta$, where $\beta$ corresponds to the average behaviour over all the users. We compute the same measure in the physical space using the mobility capacity and the new locations added to the users' \emph{activity space}, defining $R_i^{\textit{mob}} = \left<A_i\right>^{\textit{mob}} / \left<C_i\right>^{\textit{mob}}$.
Without loss in generality, we define the explorers as those with $R^{\textit{apps}}$ higher than the \nth{80} percentile and returners as those with $R^{\textit{apps}}$ lower than the \nth{20} percentile. We do similarly for the mobility.

We find that explorers adopt on average one app every 28 weeks ($\overline{A}^{\textit{apps}} = 0.72$), while keepers adopt a new app at a much slower rate ($\overline{A}^{\textit{apps}} = 0.04$). On average, mobility explorers instead embrace a new familiar location every 17 weeks ($\overline{A}^{\textit{mob}} = 1.16$), while keepers adopt a new location at a much slower rate ($\overline{A}^{\textit{mob}} = 0.11$).
The distribution of explorers and keepers are well separated in both domains, and they do not come from the same distribution~\cite{de2019strategies}.

This means that two distinct profiles of users emerge from both the physical and digital domains, which is a contribution to the existing literature in the social networks domain~\cite{miritello2013limited}. \emph{Explorers} are those who tend to try new apps and locations over time, while \emph{keepers} tend to have constant behaviour. As this exploratory behaviour is present in both the physical and digital domains, we tried to predict one from another. However, we found from our model performance that this is a very challenging task, thus posing questions whether they are two different aspects of human behaviour to be studied further.

\section{Conclusion}
Recent seminal literature has observed similarities in different aspects of human life from social interactions to mobility~\cite{alessandretti2018evidence, miritello2013limited}, from news consumption to social networks~\cite{cinelli2019selective}, but also from mobility to web browsing~\cite{barbosa2016returners, hu2018life}. In this work, we have found that many underlying mechanisms play the same role in the physical and digital worlds. These results hold even when controlling for age~\cite{de2019strategies}.
This result goes well along with recent studies on the similarity between human mobility and Web browsing~\cite{barbosa2016returners, hu2018life}, which might suggest the use of models that incorporate different aspects of human life. 

We here foresee three main avenues of future research.
First, future work might explore the possibility to learn latent user embeddings from historical app usage~\cite{chamberlain2017customer, zhao2019user}, which potentially allow the prediction of habits (next behaviour), personality, interests (e.g. gaming user profile), and social values.
Second, our results allow the study of numerous aspects of digital human behaviour. For example, future studies might describe the short and long term trends of app usage by tracking not only the apps that are discarded and adopted over time but also what kind of apps replaces the discarded ones. 
Finally, embeddings, and temporal features, along with mobility characteristics, could be jointly used to estimate different aspects of human life. For example, personality, socio-economic conditions and, in industry, the Customer lifetime value~\cite{chamberlain2017customer}).

This study does not come without limitations. We did not consider physical limitations such as battery consumption, hard disk drive space, and socio-economic factors that impact mobility.

We are increasingly converging to a world where digital devices are an integral part of our life. Thus, it is necessary to understand similarities and differences in how individuals behave online, offline and, in general, on digital devices.
We foresee the presented results will be crucial building blocks for debating a human-centric approach to AI  that builds upon the recent literature of Computational Social Science~\cite{lazer2009computational} and that creates effective models to describe, predict and simulate human behaviour.

\vspace{-0.5em}
\bibliography{ecai}
\end{document}